\documentclass[12pt]{article}
\pdfoutput=1

\usepackage{amsmath}
\usepackage{natbib}
\usepackage{graphicx,psfrag,epsf}
\usepackage{enumerate}
\usepackage{authblk}
\usepackage{hyperref}

\usepackage{titlesec}
\titlespacing*{\section}{0pt}{2mm}{0mm}
\titlespacing*{\subsection}{0pt}{2mm}{0mm}
\titlespacing*{\subsubsection}{0pt}{2mm}{0mm}
\titlespacing*{\paragraph}{0pt}{2mm}{0mm}

\def\cd#1{\texttt{#1}}
\def\cdv#1{\texttt{\begin{verbatim}{#1}\end{verbatim}}}
\def\mcd#1{\mbox{\cd{#1}}}

\newcommand{\qq}{\symbol{34}}

\begin{document}

\def\spacingset#1{\renewcommand{\baselinestretch}%
 {#1}\small\normalsize} \spacingset{1}

\title{Programming with models: writing statistical algorithms for general model structures with NIMBLE}
\author[1]{Perry de Valpine}
\author[1,2]{Daniel Turek}
\author[2]{Christopher J. Paciorek}
\author[1,2]{Clifford Anderson-Bergman}
\author[3]{Duncan Temple Lang}
\author[4]{Rastislav Bodik}
\affil[1]{\small University of California, Berkeley, Department of Environmental Science, Policy and Management}
\affil[2]{\small University of California, Berkeley, Department of Statistics}
\affil[3]{\small University of California, Davis,  Department of Statistics}
\affil[4]{\small University of California, Berkeley, Department of Electrical Engineering and Computer Science}
\date{}
\normalsize
\maketitle

\newcommand{\calculate}{\tt calculate\rm}
\newcommand{\calculateDiff}{\tt calculateDiff\rm}
\newcommand{\simulate}{\tt simulate\rm}
\newcommand{\getLogProb}{\tt getLogProb\rm}
\newcommand{\nimcopy}{\tt copy\rm}
\newcommand{\nimbleFunction}{\tt nimbleFunction\rm}
\newcommand{\assign}{\texttt{<-}}
\newcommand{\dmnorm}{\mbox{dmnorm}}
\newcommand{\ttt}{\texttt}

\bigskip
\begin{abstract}We describe NIMBLE, a system for programming statistical algorithms for general model structures within R.  NIMBLE is designed to meet three challenges: flexible model specification, a language for programming algorithms that can use different models, and a balance between high-level programmability and execution efficiency.  For model specification, NIMBLE extends the BUGS language and creates model objects, which can manipulate variables, calculate log probability values, generate simulations, and query the relationships among variables.  For algorithm programming, NIMBLE provides functions that operate with model objects using two stages of evaluation.  The first stage allows specialization of a function to a particular model and/or nodes, such as creating a Metropolis-Hastings sampler for a particular block of nodes.  The second stage allows repeated execution of computations using the results of the first stage.  To achieve efficient second-stage computation, NIMBLE compiles models and functions via C++, using the Eigen library for linear algebra, and provides the user with an interface to compiled objects.  The NIMBLE language represents a compilable domain-specific language (DSL) embedded within R. This paper provides an overview of the design and rationale for NIMBLE along with illustrative examples including importance sampling, Markov chain Monte Carlo (MCMC) and Monte Carlo expectation maximization (MCEM).
\end{abstract}

\noindent%
\vfill

\newpage
%\spacingset{1.45} 

\section{Introduction}
\label{sec:introduction}

Rapid advances in many statistical application domains are facilitated by computational methods for estimation and inference with customized hierarchical statistical models.  These include such diverse fields as ecology and evolutionary biology, education, psychology, economics, epidemiology, and political science, among others.  Although each field has different contexts, they share the statistical challenges that arise from non-independence among data -- from spatial, temporal, clustered or other sources of shared variation -- that are often modeled using unobserved (often unobservable) random variables in a hierarchical model structure \citep[e.g.,][]{banerjee-etal-03, royle-dorazio-08, cressie-wikle-11}.  

Advancement of analysis methods for such models is a major research area, including improved performance of Markov chain Monte Carlo (MCMC) algorithms \citep{brooks-etal-11}, development of maximum likelihood methods \citep[e.g.,][]{jacquier-etal-07, lele-etal-10, devalpine-12}, new approximations \citep[e.g.,][]{rue-etal-09}, methods for model selection and assessment \citep[e.g.,][]{hjort-etal-06, gelman-etal-14}, combinations of ideas such as sequential Monte Carlo and MCMC \citep{andrieu-etal-10}, and many others.  However, the current state of software for hierarchical models leaves a large gap between the limited methods available for easy application and the newer ideas that emerge constantly in the statistical literature.  In this paper we introduce a new approach to software design for programming and sharing such algorithms for general model structures, implemented in the NIMBLE package.  

The key idea of NIMBLE is to combine flexible model specification with a system for programming functions that can adapt to model structures.  This contrasts with two common statistical software designs.  In the most common approach, a package provides a fairly narrowly constrained family of models together with algorithms customized to those models.  A fundamentally different approach has been to provide a language for model specification, thereby allowing a much wider class of models.   Of these, the BUGS language \citep{gilks-etal-94} has been most widely used, with dialects implemented in WinBUGS, OpenBUGS, and JAGS \citep{lunn-etal-00, lunn-etal-12, plummer-03}.  Other tools with their own modeling language (or similar system) include AD Model Builder and its newer version, Template Model Builder \citep{fournier-etal-12,tmb}; \citet{stan}; BayesX \citep{belitz-etal-13}; and PyMC \citep{patil-etal-10}.  All of these packages have been successful for providing specific target algorithms, such as Laplace approximation and specific kinds of MCMC, but none provide a high-level way to write many different kinds of algorithms that can use the flexibly-defined models.  NIMBLE aims to do that via a compilable domain specific language \citep[DSL;][]{elliot-etal-03} embedded within R.

The design of NIMBLE uses several approaches that we think are new for statistical software.  To get started, we needed a general language for model specification, for which we adopted and extended BUGS because it has been so widely used.  NIMBLE processes BUGS code into a model object that can be used by programs: it can be queried for variable relationships and operated for simulations or probability calculations.  R was a natural fit for implementing this idea because of its high syntactic compatibility with BUGS and its ability to modify and evaluate parsed code as a first-class object, owing to its roots in Lisp and Scheme \citep{ihaka-gentleman-96}.  Second, to allow model-generic programming, we needed a way for functions to adapt to different model structures by separating one-time ``setup'' steps, such as querying a model's structure, from repeated ``run-time'' steps, such as running a Metropolis-Hastings sampler.  This was accomplished by allowing these steps to be written separately and using the concepts of specialization and staged evaluation from computer science \citep{taha-sheard-97, rompf-odersky-10}.  Third, we needed a way to allow high-level programming of algorithms yet achieve efficient computation.  This was done by creating a compiler to translate the model and run-time functions to corresponding C++ code and interfacing to the resulting objects from R.  

NIMBLE includes a domain specific language (DSL) embedded within R.  ``Run-time'' code can be thought of as a subset of R with some special functions for handling models.  Programming in NIMBLE is a lot like programming in R, but the DSL formally represents a distinct language defined by what is allowed for compilation.  NIMBLE stands for Numerical Inference for statistical Models using Bayesian and Likelihood Estimation.  

The rest of this paper is organized as follows.  First we give an overview of NIMBLE's motivation and design, discussing each major part and how they interact, without specific implementation details.  Then we give examples of three algorithms -- importance sampling, MCMC, and Monte Carlo expectation maximization \citep[MCEM;][]{wei-tanner-90, levine-casella-01} -- operating on one model to illustrate how the pieces fit together to provide a flexible system.  These examples, and the more complete code available in the online supplement, provide an introduction to NIMBLE's implementation.  A complete user manual is available at the project web site (R-nimble.org).

\section{Overview of NIMBLE}
\label{sec:overview-of-NIMBLE}

NIMBLE comprises three main components (Fig. 1): a new implementation of BUGS (with extensions) as a model declaration language (Fig. 1:A-C); the \texttt{nimbleFunction} system for programming with models (Fig. 1:D-G); and the NIMBLE compiler for \cd{model} objects and \texttt{nimbleFunction}s (Fig. 1:H-J).

\begin{figure}[!ht]
\begin{center}
\includegraphics{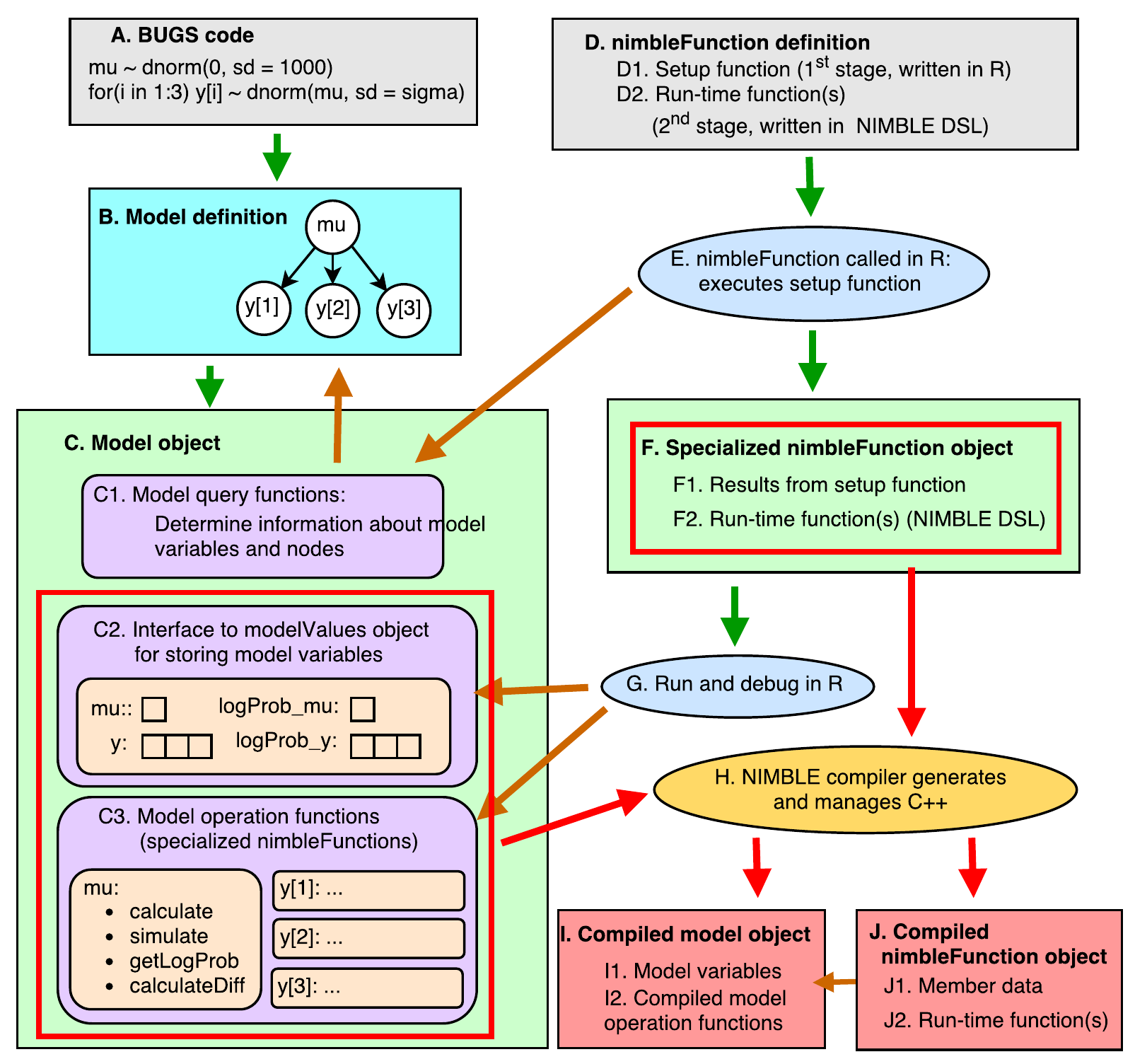}
\caption{\linespread{1}\selectfont{} Overview of NIMBLE.  Left side: A model starts as BUGS code (A), which is turned into a \cd{model definition} object (B), which creates an uncompiled \cd{model} object (C).  Right side: A \cd{nimbleFunction} starts as model-generic code (D). It is specialized to a model and/other arguments by executing its \cd{setup} function (E), which may inspect the \cd{model} structure (brown arrow, using C1).  This returns an uncompiled, specialized \cd{nimbleFunction} object (F).  Its run-time function(s) can be executed in R, using the uncompiled model (brown arrows), to debug algorithm logic (G).  Parts of the \cd{model} and \cd{nimbleFunction} (red boxes) can be compiled (H), creating objects (I, J) that can be used from R similarly to their uncompiled counterparts.
 Gray = code.  Blue = R execution.  Green, purple \& tan = Uncompiled objects that run in pure R.  Green arrows = pre-compilation workflow.  Red boxes \& arrows  = compilation workflow. }
\end{center}
\end{figure}

\subsection{Design rationale}
\label{rationale}
\label{sec:progr-with-models}

The goal of NIMBLE is to make it easier to implement and apply a variety of algorithms to any model defined as a directed acyclic graph (DAG).  For example, we might want to use (i) several varieties of MCMC to see which is most efficient \citep{brooks-etal-11}, including programmatic exploration of valid MCMC samplers for a particular model; (ii) other Monte Carlo methods such as sequential Monte Carlo \citep[SMC, a.k.a. ``particle filters'';][]{doucet-etal-01} or importance sampling; (iii) modular combinations of methods, such as combination of particle filters and MCMC in state-space time-series models \citep{andrieu-etal-10} or combination of Laplace approximation and MCMC for different levels of the model; (iv) algorithms for maximum likelihood estimation such as MCEM and data cloning \citep{lele-etal-07, jacquier-etal-07, devalpine-12}; (v) methods for model criticism, model selection, and estimation of prediction error \citep{vehtari-ojanen-12} such as Bayesian cross-validation \citep{gelfand-etal-92, stern-cressie-00}, calibrated posterior predictive p-values \citep{hjort-etal-06} or alternatives to DIC such as WAIC \citep{watanabe-10, spiegelhalter-etal-14}; (vi) ``likelihood free'' or ``plug-and-play'' methods such as synthetic likelihood \citep{wood-10}, approximate Bayesian computation \citep[ABC;][]{marjoram-etal-03}, or iterated filtering \citep{ionides-etal-06}; (vii) parametric bootstrapping of any of the above ideas; or (viii) the same model and algorithm for multiple data sets. These are just some of many ideas that could be listed.  

There are several reasons the above kinds of methods have been difficult to handle in general software. First, if one has wanted to write a package providing a new general method, one has had to ``reinvent the wheel'' of model specification.  This means deciding on a class of allowed models, writing a system for specifying the models, and writing the algorithm to use that system.  Creating model specification systems for each package is difficult and tangential to the statistical algorithms themselves.  It also results in multiple different systems for specifying similar classes of models.  For example, lme4 \citep{lme4}, MCMCglmm \citep{hadfield-10}, R-INLA \citep{martins-etal-13}, and others each use a different system for GLMM specification.  We desired a system with the flexibility of BUGS for declaring a wide range of models, while allowing different algorithms to use the same representation of a given model.

A second limitation of current designs arises from the tension between expressing algorithms easily in a high-level language and obtaining good computational performance.  High-level languages, especially R, can be slow, but low-level languages like C++ require much greater implementation effort and customization to different problems.  A common solution to this problem has been to write computationally intensive steps in a low-level language and call them from the high-level language.  This results in code that is less general and less accessible to other developers.  Most of the general MCMC packages represent an extreme case of this phenomenon, with the algorithms hidden in a ``black box'' unless one digs into the low-level code.  We wanted to keep more programming in a high-level language and use compilation to achieve efficiency.

\subsection{Specifying models: Extending the BUGS language}
\label{sec:adopt-bugs-lang}

We chose to build upon the BUGS language because it has been widely adopted \citep{lunn-etal-09}.  Many books use BUGS to teach Bayesian statistical modeling \citep[e.g.,][]{lancaster-04, kery-schaub-11, vidakovic-11}, and domain scientists find that it helps them to reason clearly about models \citep{kery-schaub-11}.  Many users of the BUGS packages think of BUGS as nearly synonymous with MCMC, but we distinguish BUGS as a DSL for model specification from its use in MCMC packages.  The differences between BUGS dialects in JAGS, OpenBUGS, and WinBUGS are not important for this paper. 

\subsubsection{BUGS, model definitions, and models}

When NIMBLE processes BUGS code (Fig. 1:A), it extracts all semantic relationships in model declarations and builds two primary objects from them.  The first is a \texttt{model definition} object (Fig. 1:B), which includes a representation of graph nodes (also called \textit{vertices} in graph theory) and edges.  The second is a \texttt{model} object (Fig. 1:C), which contains functions for investigating model structure (Fig. 1:C1), objects to store values of model variables (Fig. 1:C2) and sets of functions for model calculations and simulations (Fig 1:C3).  One \texttt{model definition} can create multiple \texttt{model}s with identical structure.  Normally a user interacts only with the \texttt{model} object, which may use its \texttt{model definition} object internally (Fig. 1: brown arrow from C1 to B).

At this point, it will be useful to introduce several concepts of \cd{model definition} objects and \cd{model} objects designed to accommodate the flexibility of BUGS. Each BUGS declaration creates a \textit{node}, which may be stochastic or deterministic (``logical'' in BUGS). For example, node \cd{y[3]} may be declared to follow a normal distribution with mean \cd{mu} and standard deviation \cd{sigma} (Fig 1:A). That would make \cd{mu} and \cd{sigma} \textit{parents} of \cd{y[3]} and \cd{y[3]} a \textit{dependent} (or \textit{dependency}) of \cd{mu} and \cd{sigma}.  NIMBLE uses \textit{variable} to refer to a possibly multivariate object whose elements represent one or more nodes.  For example, the variable \cd{y} includes all nodes declared for one or more elements of \cd{y} (e.g., \cd{y[1], y[2], y[3]}). A node can be multivariate, and such nodes can be occur arbitrarily in contiguous scalar elements of a variable.  Groups of nodes in a variable may be declared by iteration, such that their role in the model follows a pattern, but they may also be declared separately, so it cannot be assumed in later processing that they do follow a pattern.  The \cd{model definition} uses abstractions for variables, nodes, and their graph relationships that supports handling of interesting cases.  For example, a program may need to determine the dependencies of just one element of a multivariate node, even though that element is not itself a node.

Processing BUGS code in a high-level language like R facilitates some natural extensions to BUGS.  First, NIMBLE makes BUGS extensible by allowing new functions and distributions to be provided as \nimbleFunction s.  Second, NIMBLE can transform a declared graph into different, equivalent graphs that may be needed for different implementation contexts.  For example, NIMBLE implements alternative parameterizations for distributions by automatically inserting nodes into the graph to transform from one parameterization to another.  If the function that ultimately executes gamma probability density calculations needs the \cd{rate} parameter but the BUGS code declares a node to follow a gamma distribution using the \cd{scale} parameter (related by rate = 1/scale), a new node is inserted to calculate \cd{1/scale}, which is then used as the needed gamma \cd{rate}.  If any other declaration invokes the same reparameterization, it will use the same new node.  Another important, optional, graph transformation occurs when the parameter of a distribution is an expression.  In that case a separate node can be created for the expression's value and inserted for use where needed.  This is useful when an algorithm needs access to the value of a parameter for a particular node, such as for conjugate distribution relationships used in Gibbs sampling and other contexts. A third extension is that BUGS code can be used to define a set of alternative models by including conditional statements (i.e., \cd{if-then-else}) that NIMBLE evaluates (in R) when the \cd{model definition} is created.  This avoids the need to copy and modify entire BUGS model definitions for each alternative model, the standard practice when using previous BUGS packages.

NIMBLE uses a more general concept of \textit{data} than previous BUGS packages.  In previous packages, a model cannot be defined without its data.  In NIMBLE, \textit{data} is a label for the role played by certain nodes in a model.  For example, nodes labeled as data are excluded from calls to simulate new values into the model by default, to avoid over-writing observed values, but this default can be over-ridden by a programmer who wishes to simulate fake data sets from the model.  The data label is distinct from the actual \textit{values} of nodes labeled as data, which can be programmatically changed. For example, one might want to iterate over multiple data sets, inserting each one into the data nodes of a model and running an algorithm of interest for each.

At the time of this writing, some BUGS features are not implemented.  Most notably, NIMBLE does not yet allow stochastic indexing, i.e., indices that are not constants.

\subsubsection{How \cd{model} objects are used}

A \texttt{model} object is used in two ways from R and/or \texttt{nimbleFunction}s.  First, one may need to query node relationships,  a common step in setup code (Fig 1: brown arrow from E to C1).  For example, consider a \texttt{nimbleFunction} for a Metropolis-Hastings MCMC sampler (shown in detail later).  In one instance, it may be needed to sample a node called \cd{mu[2]}, in another to sample a node called \cd{x[3, 5]}, and so on.  We refer to the node to be sampled as the \textit{target node}.  The setup stage of the \texttt{nimbleFunction} can query the \texttt{model} object to determine what stochastic nodes depend on the target node and save that information for repeated use by run-time code.  Or it may be that an R function needs to query a \texttt{model} object, for example to determine if it conforms to the requirements for a particular algorithm.  The implementation of the \texttt{model} uses its \texttt{model definition} to respond to such queries, but the  \texttt{nimbleFunction} programmer is protected from that detail.

Other examples of model queries include determining:

\begin{itemize}
\itemsep=-3pt
\item Topologically sorted order of nodes, which means an order valid for sequential calculations or simulations.
\item All nodes or variables in the model of a particular type, such as stochastic, deterministic, and/or data nodes.
\item The position of nodes in the model: e.g., \textit{top} nodes have no stochastic parents; \textit{end} nodes have no stochastic dependents; and \textit{latent} nodes have stochastic parents and dependents. 
\item The nodes contained in an arbitrary subset of variable elements.  For example, \cd{x[3:5]} may represent the three scalar nodes \cd{x[3]}, \cd{x[4]}, and \cd{x[5]}, or it may represent one scalar node \cd{x[3]} and one multivariate node \cd{x[4:5]}, or other such combinations.
\item Nodes or expressions with certain semantic relationships, such as the node or expression for the \cd{rate} parameter of a gamma distribution.
\item A variety of kinds of dependencies from a set of nodes.  For example, stochastic dependencies (also called ``Markov blankets'') include all paths through the graph terminating at, and including, stochastic nodes.  These are needed for many algorithms.  In other cases, stochastic dependencies without data nodes are needed, such as for one time-step of a particle filter.  Deterministic dependencies are like stochastic dependencies but omit the stochastic nodes themselves.  This kind of dependency is useful following the assignment of a value to a node to ensure descendent stochastic nodes use updated parameter values.
\end{itemize}

The second way \texttt{model} objects are used is to manage node values and calculations, both of which are commonly needed in run-time functions (Fig 1:F2).  A \cd{model} object contains each model variable and any associated log probabilities (Fig 1:C2).  It also can access functions for calculating log probabilities and generating simulations for each node (Fig 1:C3). These functions are constructed as \nimbleFunction s from each line of BUGS code.  Specifically, each node has a \nimbleFunction\ with four run-time functions:
\begin{itemize}
\itemsep=-3pt
\item \calculate: For a stochastic node, this calculates the log probability mass or density function, stores the result in an element of the corresponding log probability variable (Fig. 1: C2), and returns it.  For a deterministic node, \calculate\ executes its computation, stores the result as the value of the node, and returns 0.
\item \cd{calculateDiff}: This is like \calculate\ except that for a stochastic node it returns the difference between the new log probability value and the previously stored value. This is useful for iterative algorithms such as Metropolis-Hastings-based MCMC.
\item \simulate: For a stochastic node, this generates a draw from the distribution and stores it as the value of the node.  \simulate\ has no return value.  For a deterministic node, \simulate\ is identical to \calculate\ except that it has no return value.
\item \getLogProb: For a stochastic node, this returns the currently stored log probability value corresponding to the node.  For a deterministic node, this returns 0.
\end{itemize}
A \texttt{model} object has functions of the same names to call each of these node functions for an ordered sequence of nodes.  With the exception of \simulate, these return the sum of the values returned by the corresponding node functions (e.g., the sum of log probabilities for \cd{calculate}).  A typical idiom for model-generic programming is to determine a vector of nodes by inspecting the model in setup code and then use it for the above operations in run-time code.

\subsubsection{\cd{modelValues} objects for storing multiple sets of model values}
\label{sec:model-valu-modelv}
 A common need for hierarchical model algorithms is to store multiple sets of values for multiple model variables, possibly including their associated log probability variables.  NIMBLE provides a \ttt{modelValues} data structure for this purpose.  When a \cd{model definition} is created, it builds a specification for the related \cd{modelValues} class.   When a \cd{model} object is created, it includes an object of the \cd{modelValues} class as a default location for model values.  New \cd{modelValues} classes and objects can be created with whatever variables and types are needed. Examples of uses of \cd{modelValues} objects are: storing the output of MCMC; storing a set of simulated node values for input to importance sampling; and storing a set of ``particle'' values and associated log probabilities for a particle filter.  

\subsection{Programming with models}

One can use model objects arbitrarily in R, but NIMBLE's system for model-generic programming is based on \texttt{nimbleFunction}s (Fig 1:D).  Separate function definitions for the two evaluation stages -- one \cd{setup} function and one or more run-time functions -- are written within the \texttt{nimbleFunction} (Fig 1:D1,D2). The purpose of a \cd{setup} function is to \textit{specialize} a \texttt{nimbleFunction} to a particular \cd{model} object, nodes, or whatever other arguments are taken by the \cd{setup} function.  This typically involves one-time creation of objects that can be used repeatedly in run-time code.  Such objects could be results from querying the model about node relationships, specializations of other \nimbleFunction s, new \cd{modelValues} objects, or results from arbitrary R code.  When a \texttt{nimbleFunction} is called, the arguments are passed to the \cd{setup} function, which is evaluated in R (Fig 1:E).  The \nimbleFunction\ saves the evaluation environment and creates the return object.  The return object is an instance of a custom-generated class whose member functions are the run-time function(s) (Fig 1:F).  

The two-stage evaluation of \texttt{nimbleFunction}s is similar to a function object (functor) system: the \texttt{nimbleFunction} is like an implicit class definition, and calling it is like instantiating an object of the class with initialization steps done by the \cd{setup} function.  However the \nimbleFunction\ takes care of steps such as determining which objects created during \cd{setup} evaluation need to become member data in a corresponding class definition and determining their types from specialized instances of the \nimbleFunction.  As a result, the programmer can focus on higher level logic. 

The run-time functions include a default-named \cd{run} function and arbitrary others.  These are written in the NIMBLE DSL, which allows them to be evaluated natively in R (Fig 1:G) or compiled into C++ class methods (Fig 1:H).  The former allows easier debugging of algorithm logic, while the latter allows much faster execution.  It is also possible to omit the \cd{setup} function and provide a single \cd{run} function, which yields a simple function in the NIMBLE DSL that can be compiled to C++ but has no first-stage evaluation and hence no specialization.  For both \cd{model}s and \nimbleFunction s, the R objects that use compiled or uncompiled versions provide a largely identical interface to the R user. 

 The NIMBLE DSL supports control of \texttt{model} and \texttt{modelValues} objects, common math operations, and basic flow control.  Control of \texttt{model} objects includes accessing values of nodes and variables as well as calling \calculate, \calculateDiff, \simulate, and \getLogProb\ for vectors of nodes.  With these basic tools, a run-time function can \textit{operate} a model: get or set values, simulate values, and control log probability calculations.   Use of \texttt{modelValues} objects includes setting and accessing specific values and copying arbitrary groups of values between \texttt{model} and/or other \texttt{modelValues} objects using the special \nimcopy\ operation.  Together these uses of \texttt{modelValues} facilitate iteration over sets of values for use in a \texttt{model} object.  For example, a \texttt{modelValues} object might contain the ``particle'' sample of a particle filter, and the run function could iterate over them, using each one in the model for some simulation or calculation. Supported math operations include basic (vectorized) math, linear algebra, and probability distribution calculations. 

The two-stage evaluation system works naturally when one \nimbleFunction\ needs to use other \nimbleFunction s.  One \nimbleFunction\ can specialize another \nimbleFunction\ in its \cd{setup} code, or take it as a \cd{setup} argument, and then use it in run-time code.  In addition one can create vectors of \nimbleFunction s.  There is a simple \nimbleFunction\ class inheritance system that allows labeling of different \nimbleFunction s that have the same run-time function prototype(s).  For example, a \cd{nimbleFunction} for MCMC contains a vector of \nimbleFunction s, each of which updates (samples) some subset of the model.  The latter \nimbleFunction s inherit from the same base class.  This is a light burden for the NIMBLE programmer and allows the NIMBLE compiler to easily generate a simple C++ class hierarchy.  One can also create numeric objects, lists of same-type numeric objects, and customized \cd{modelValues} objects in \cd{setup} code for use in run-time code.

The \texttt{nimbleFunction} system is designed to look and feel like R in many ways, but there are important differences.  The \cd{setup} function does not have a programmer-defined return value because the \texttt{nimbleFunction} system takes charge by returning a specialized \texttt{nimbleFunction} (ready for run-time function execution) after calling its \cd{setup} function.  More importantly, the run-time function(s) have some highly non-R-like behavior.  For efficient C++ performance, they pass arguments by reference, opposite to R's call-by-value semantics.  To support the static typing of C++, once an object name is used it cannot subsequently be assigned to a different-type object.  And type declarations of arguments and the return value are required in order to simplify compiler implementation.  To a large extent, other types are inferred from the code.

\subsection{The NIMBLE compiler}
\label{sec:comp-from-nimble}
A thorough description of the NIMBLE compiler is beyond the scope of this paper, but we provide a brief overview of how \nimbleFunction s and \cd{model}s are mapped to C++ and how NIMBLE manages the use of the compiled C++.  The NIMBLE compiler generates a C++ class definition for a \nimbleFunction.  Results of \cd{setup} code that are used in run-time code are turned into member data.  The default-named \cd{run} member function and other explicitly defined run-time functions are turned into C++ member functions.  Once the C++ code is generated, NIMBLE calls the C++ compiler and loads the resulting shared object into R.  Finally, NIMBLE dynamically generates an R reference class definition to provide an interface (using active bindings) to all member data and functions of objects instantiated from compiled C++ (Fig 1:J).  This creates an object with identical interface (member functions) as its uncompiled counterpart for the R user.  When there are multiple instances (specializations) of the same \cd{nimbleFunction}, they are built as multiple objects of the same C++ class.  If a \cd{nimbleFunction} is defined with no \cd{setup} code, then there is no first-stage evaluation, and the corresponding C++ is a function rather than a class.

Compilation of \cd{model}s involves two components.  Each line of BUGS code is represented as a custom-generated \cd{nimbleFunction} with \calculate, \cd{calculateDiff}, \simulate, and \getLogProb\ run-time functions.  These are compiled like any other \nimbleFunction, the only difference being inheritance from a common base class.  This facilitates NIMBLE's introduction of extensibility for BUGS by allowing new functions and distributions to be provided as \nimbleFunction s.  The variables of a \cd{model} and \cd{modelValues} are implemented by generating simple C++ classes with appropriate member objects.  Like \nimbleFunction s, both \cd{model}s and \cd{modelValue}s objects are automatically interfaced via R objects that have similar interfaces to their uncompiled counterparts (Fig 1:I).  

For the most part, the compiler infers types and dimensionality of numeric variables and generates code for run-time size-checking and resizing.  The exceptions include required declaration of run-time argument types and the return type as well as situations where size inference is not easy to implement.  NIMBLE includes a library of functions and classes used in generated C++.  Vectorized math and linear algebra are implemented by generating code for the Eigen C++ library \citep*{eigenweb}.  Basic \cd{for}-loops for numeric iterators and basic flow control using \cd{if-then-else} and \cd{do-while} constructs are supported.  The actual compilation processing converts run-time code into an abstract syntax tree (AST) with an associated symbol table, which are annotated and transformed into a C++ syntax tree.  A set of R classes for representing C++ code was developed for this purpose.

Compilation of \nimbleFunction s harnesses completed first-stage evaluation (specialization) in several ways.  First, contents of objects created during \cd{setup} evaluation can be directly inspected to determine types.  Second, the compiler uses \textit{partial evaluation} to simplify the C++ code and types needed.  For example, the compiler resolves nodes in \cd{model} objects at compile time so that the C++ code can find the right object by simple pointer dereferencing.  It also converts vectors of nodes into different kinds of objects depending on how they are used in run-time code.  Such partial evaluation is done in the \cd{setup} environment, essentially as a compiler-generated extension to the \cd{setup} code.

\section{Examples}
In this section we present some examples of model-generic programming and the algorithm composition it supports.  This section includes more implementation details, including some code for discussion.  Specifically, we show how importance sampling and Metropolis-Hastings sampling are implemented as \nimbleFunction s.  Then we show how an MCMC is composed of multiple samplers that can be modified programmatically from R.  Finally we show an example of composing an algorithm that uses MCMC as one component, for which we choose MCEM.  Complete code to replicate the examples is provided in the supplement.

As a model for illustration of these algorithms, we choose the pump model from the WinBUGS/OpenBUGS suite of examples \citep{lunn-etal-12} because it is simple to explain and use. We assume some familiarity with BUGS. The BUGS code is:
\begin{verbatim}
pumpCode <- nimbleCode({ 
  for (i in 1:N){
      theta[i] ~ dgamma(shape = alpha, rate = beta) ## random effects
      lambda[i] <- theta[i]*t[i]     ## t[i] is explanatory data
      x[i] ~ dpois(lambda[i])        ## x[i] is response data
  }
  alpha ~ dexp(1.0)                  ## priors for alpha and beta
  beta ~ dgamma(0.1, 1.0)
})
\end{verbatim}
Here \cd{x} and \cd{t} are to be provided as data (not shown), \cd{theta} are random effects, and \cd{alpha} and \cd{beta} are the parameters of interest.   We have written the gamma distribution for \cd{theta[i]} using named parameters to illustrate this extension of BUGS.  Creation of a model object called \cd{pumpModel} from the \cd{pumpCode} is shown in the supplement.

\subsection{Importance sampling}

Importance sampling is a method for approximating an expected value from a Monte Carlo sample \citep{givens-hoeting-12}. It illustrates the glaring gap between algorithms and software: although it is an old and simple idea, it is not easily available for general model structures.  It involves sampling from one distribution and weighting each value so the weighted sample represents the distribution involved in the expected value.  It can be used to approximate a normalizing constant such as a likelihood or Bayes factor.  (It can also be combined with a resampling step to sample from a Bayesian posterior, i.e., Sampling/Importance Resampling.)  

\newcommand{\mIS}{\mbox{\scriptsize{IS}}}
\newcommand{\mlam}{\mbox{\cd{lambda}}}
\newcommand{\my}{\mbox{y}}

For the pump model, suppose we want to use importance sampling to approximate the marginal likelihood of \cd{x[1:3]}, which requires integrating over the first three random effects, \cd{theta[1:3]}, given values of \cd{alpha} and \cd{beta}.  This is an arbitrary subset of the model for illustration.  To do so one simulates a sample $\mcd{theta[1:3]}_k \sim P_{\mIS}(\mcd{theta[1:3]})$, $k = 1\ldots m$, where $P_{\mIS}$ is a known distribution.  For mathematical notation, we are mixing the code's variable names with subscripts, so that $\cd{theta[1:3]}_k$ is the $k^{\mbox{th}}$ simulated value of \cd{theta[1:3]}.  Then the likelihood is approximated as 
\begin{equation}
\label{ISeq}
P(\mcd{x[1:3]})  \approx \frac{1}{m} \sum_{k = 1}^m P(\mcd{x[1:3]} | \mcd{theta[1:3]}_k) \frac{P(\mcd{theta[1:3]}_k)}{P_{\mIS}(\mcd{theta[1:3]}_k)}
\end{equation}
where $P(\cdot)$ indicates the part of the model's probability density or mass labeled by its argument.  The ratio on the right is the importance weight for the $k^{\mbox{th}}$ value of \cd{theta[1:3]}.  To keep the example concise, we assume the programmer already has a function (in R or NIMBLE) to sample from $P_{\mIS}$ and calculate the denominator of the weights.  Our example shows the use of NIMBLE to calculate (\ref{ISeq}) from those inputs. 

Model-generic NIMBLE code for calculation of (\ref{ISeq}) is as follows:
\begin{verbatim}
importanceSample <- nimbleFunction(
    setup = function(model, sampleNodes, mvSample) {
        calculationNodes <- model$getDependencies(sampleNodes)
    },
    run = function(simulatedLogProbs = double(1)) {
        ans <- 0.0                                             # (1)
        for(k in 1:getsize(mvSample)) {                        # (2)
            copy(from = mvSample, to = model,                  # (3)
                 nodes = sampleNodes, row = k)
            logProbModel <- model$calculate(calculationNodes)  # (4)
            if(!is.nan(logProbModel))                          # (5)
              ans <- ans + exp(logProbModel - simulatedLogProbs[k])
        }
        return(ans/getsize(mvSample))                          # (6)
        returnType(double(0))                                  # (7)
    }
)
\end{verbatim}
The specialization step for our example would be \cd{pumpIS <- importanceSample(model = pumpModel, sampleNodes = "theta[1:3]", mvSample = ISsample)}.  Note that the arguments to \cd{importanceSample} are defined in its \cd{setup} function.  \cd{model} is given as the \cd{pumpModel} object created above.  \cd{sampleNodes} -- the set of nodes over which we want to integrate by importance sampling -- is provided as a character vector using R's standard variable subset notation.  The \cd{ISsample} object passed as the \cd{mvSample} argument is a \cd{modelValues} object for providing the values sampled from $P_{\mIS}$. It does not need to be populated with sample values until the \cd{run} function is called.  Rather, at the \cd{setup} stage, it just binds \cd{mvSample} to (a reference to) \cd{ISsample} for use in the \cd{run} code.  

The only processing done in the \cd{setup} code is to query the model for the vector of ordered stochastic dependencies of the \cd{sampleNodes}.  These are needed in the \cd{run} code to calculate the necessary part of the model in topologically sorted order.  The model is queried using \cd{getDependencies}, and the result saved in \cd{calculationNodes}.  In this case, \cd{calculationNodes} will turn out to be (\cd{theta[1], theta[2], theta[3], lambda[1], lambda[2], lambda[3], x[1], x[2]}, \cd{x[3]}).  This means that \cd{model\$calculate(calculationNodes)} will return $\log \left( P(\mcd{x[1:3]} | \mcd{theta[1:3]}_k) P(\mcd{theta[1:3]}_k) \right)$. 

The \cd{run} code illustrates several features of the NIMBLE DSL.  It shows type declaration of the \cd{simulatedLogProbs} argument as a vector of doubles (double-precision numbers) and (7) the return type as a scalar double.  \cd{simulatedLogProbs}  represents the vector of $P_{\mIS}(\mcd{theta[1:3]}_k)$ values.  In another implementation of importance sampling, this could be included in the \cd{mvSample} object, but we use it here to illustrate a run-time argument.  The body of the \cd{run} function (1) initializes the answer to zero; (2) iterates over the samples in \cd{mvSample}; (3) copies values of the \cd{sampleNodes}  from \cd{mvSample} into the model; (4) \calculate s the sum of log probabilities of \cd{calculationNodes}; and (5) uses basic \cd{if-then} logic and math to accumulate the results. 

The most important insight about \cd{importanceSample} is that it is model-generic: nothing in the \cd{setup} code or \cd{run} code is specific to the pump model or nodes \cd{theta[1:3]}. 

\subsection{Metropolis-Hastings samplers}
\label{sec:writ-funct-that}
Next we illustrate a Metropolis-Hastings sampler with a normally-distributed random-walk proposal distribution.
The model-generic code for this is:

\begin{verbatim}
simple_MH <- nimbleFunction(
  setup = function(model, currentState, targetNode) {
    calculationNodes  <- model$getDependencies(targetNode)
  },
  run = function(scale = double(0)) {
    logProb_current <- model$getLogProb(calculationNodes)               # (1)
    proposalValue <- rnorm(1, mean = model[[targetNode]], sd = scale)   # (2)
    model[[targetNode]] <<- proposalValue                               # (3)
    logProb_proposal <- model$calculate(calculationNodes)               # (4)
    log_Metropolis_Hastings_ratio <- logProb_proposal - logProb_current # (5) 
    accept <- decide(log_Metropolis_Hastings_ratio)                     # (6)
    if(accept)
      copy(from = model, to = currentState, row = 1,                   # (7a)
           nodes = calculationNodes, logProb = TRUE)
    else
      copy(from = currentState, to = model, row = 1,                   # (7b)
           nodes = calculationNodes, logProb = TRUE)
    return(accept)
    returnType(integer(0))
  })
\end{verbatim}
Suppose we want a sampler for \cd{theta[4]} in the pump model.  An example specialization step would be \cd{theta4sampler <- simple\_MH(model = pumpModel, currentState = mvState, targetNode = "theta[4]")}.  Here \cd{mvState} is a \cd{modelValues} object with variables that match those in the model, with only one of each.  This is used to store the current state of the \cd{model}.  We assume that on entry to the \cd{run} function, \cd{mvState} will contain a copy of all model variables and log probabilities, and on exit the \cd{run} function must ensure that the same is true, reflecting any updates to those states.   As in the importance sampling example, the only real work to be done in the \cd{setup} function is to query the model to determine the stochastic dependencies of the \cd{targetNode}.  In this case \cd{calculationNodes} will be \cd{(theta[4], lambda[4], x[4])}.

The \cd{run} function illustrates the compactness of expressing a Metropolis-Hastings algorithm using language elements like \calculate, \getLogProb, \cd{copy}, and list-like access to a model node.  The \cd{scale} run-time argument is the standard deviation for the normally distributed proposal value.  In the full, released version of this algorithm (\cd{sampler\_RW}), the \cd{setup} code includes some error trapping, and there is additional code to implement adaptation of the \cd{scale} parameter \citep{haario-etal-01} rather than taking it as a run-time argument.  The simplified version here is less cluttered for illustration.  In addition the full version is more efficient by using \calculateDiff\ instead of both \getLogProb\ and \calculate, but here we use the latter to illustrate the steps more clearly.

The lines of \cd{run} (1) obtain the current sum of log probabilities of the stochastic dependents of the target node (including itself); (2) simulate a new value centered on the current value (\cd{model[[targetNode]]}); (3) put that value in the model; (4) calculate the new sum of log probabilities of the same stochastic dependents; (5) determine the log acceptance probability; (6) call the utility function \cd{decide} that determines the accept/reject decision; and (7) copy from the \cd{model} to the \cd{currentState} for (7a)  an acceptance or (7b) vice-versa for a rejection.  Again, the \cd{setup} and \cd{run} functions are fully model-generic.

This example illustrates natural R-like access to nodes and variables in models, such as \cd{model[[targetNode]]}, but making this model-generic leads to some surprising syntax.  Every node has a unique character name that includes indices, such as \cd{{\qq}theta[4]{\qq}}.  This leads to the syntax \cd{model[[{\qq}theta[4]{\qq}]]}, rather than \cd{model[[{\qq}theta{\qq}]][4]}.  The latter is also valid, but it is not model-generic because, in another specialization of \cd{simple\_MH}, \cd{targetNode} may have a different number of indices.  For example, if \cd{targetNode} is {\cd{{\qq}y[2, 3]{\qq}}, \cd{model[[targetNode]]} accesses the same value as \cd{model[[{\qq}y{\qq}]][2,3]}.  The NIMBLE DSL also provides vectorized access to groups of nodes and/or variables.

\subsection{MCMC}
To illustrate a full set of MCMC samplers for a model, we do not provide \cd{nimbleFunction} code as above but rather illustrate the flexibility provided by managing sampler choices from R.   The first step in creating an MCMC is to inspect the model structure to decide what kind of sampler should be used for each node or block of nodes.  An R function (\cd{configureMCMC}) does this and returns an object with sampler assignments, which can be modified before creating the \nimbleFunction s to execute the MCMC.  Since this is all written in R, one can control its behavior, modify the code, or write a completely new MCMC system.  Once the user is happy with the MCMC configuration, the corresponding suite of specialized \nimbleFunction s can be built, compiled, and executed.

In the case of the pump model (see supplement), we choose for illustration to start with normal adaptive random walk samplers rather than Gibbs samplers.
 It is apparent from Figure 2 (left panel) that the posterior is correlated between \cd{alpha} and \cd{beta}. One might then customize the sampler choices using this knowledge.  For example, one can insert a bivariate (block) adaptive random walk sampler and then re-compile the MCMC.  This results in improved mixing, reflected as lower autocorrelations of the chain (Fig. 2, middle panel) and higher effective sample size per second of computation (Fig. 2, right panel).

\begin{figure}[h]
\begin{center}
\includegraphics{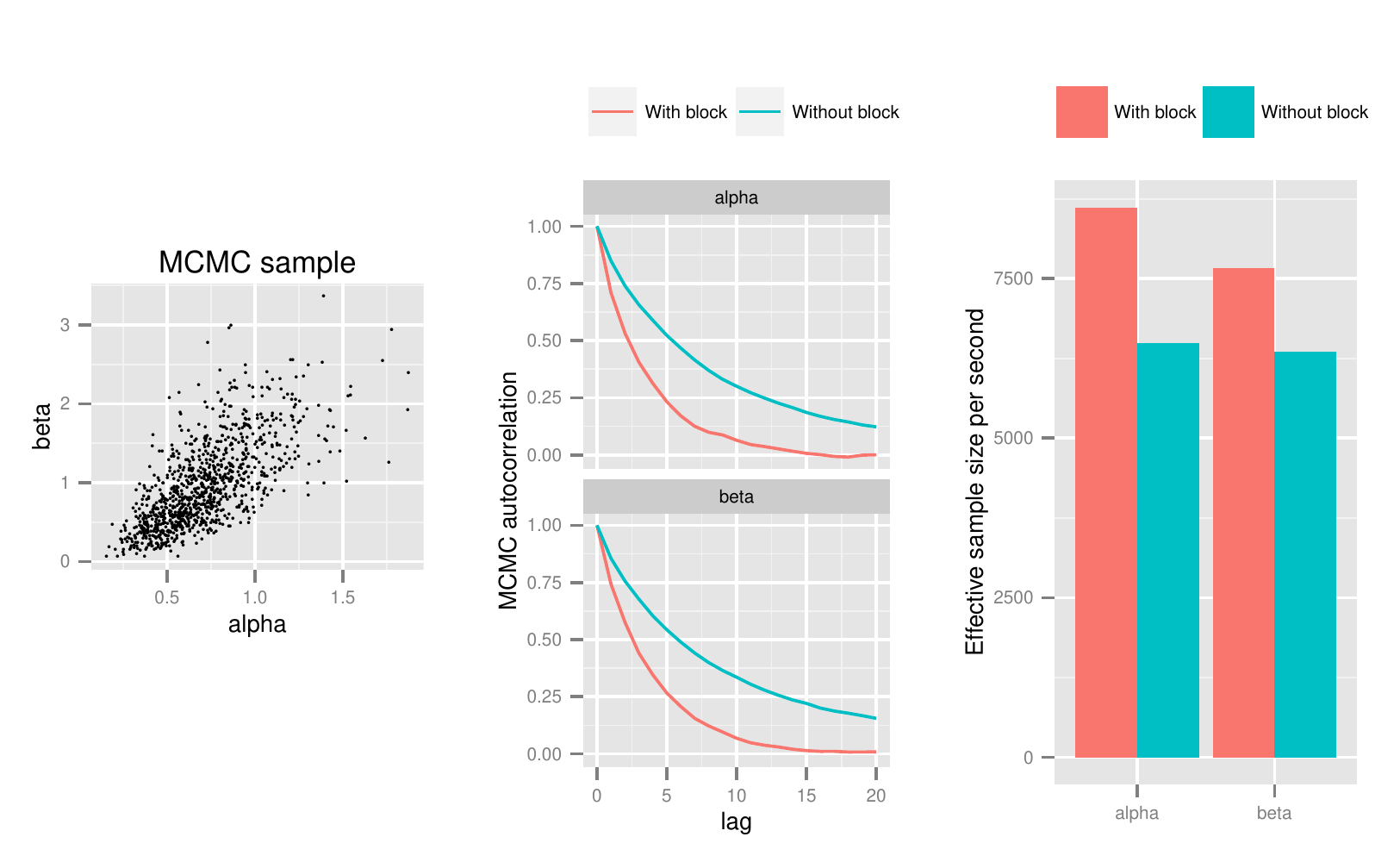}
\caption{\linespread{1}\selectfont{} Example of how high-level programmability and compilation allow flexible composition of efficient algorithms.  This uses the ``pump'' model from the classic BUGS examples.  Left panel: Parameters $\alpha$ and $\beta$ show posterior correlation.  Middle panel: MCMC mixing is summarized using the estimated autocorrelation function.  When a bivariate (block) adaptive random walk sampler is added to the suite of univariate adaptive random walk samplers, the chain autocorrelation decreases, reflecting better mixing.  Right panel: Computational performance measured as the effective sample size per second of computation time is greater with the block sampler included.}
\end{center}
\end{figure}

\subsection{Monte Carlo Expectation Maximization}
\label{sec:monte-carlo-expect}

MCEM is a widely known algorithm for maximum likelihood estimation for hierarchical models.  It is used instead of the EM algorithm when the ``expectation'' step cannot be determined analytically.
To our knowledge, there has been no previous implementation of MCEM that can automatically be applied to the range of model structures provided by BUGS. MCEM works by iterating over two steps: (1) MCMC sampling of the latent states given fixed parameters (top-level nodes); and (2) optimization with respect to (non-latent) parameters of the average log probability of the MCMC sample. NIMBLE provides a \cd{buildMCEM} function in which step (1) is implemented by creating an MCMC configuration with samplers only for latent states, and step (2) is implemented by calling one of R's optimizers with a compiled \nimbleFunction\ as the objective function.  The top level of control of the algorithm is an R function that alternates between these steps.  For the pump model, the MCEM quickly settled within 0.01 of the published values of 0.82 and 1.26 for \cd{alpha} and \cd{beta} \citep{george-etal-93}, which we consider to be within Monte Carlo error.

\section{Discussion}
\label{sec:discussion}
We have introduced a system for combining a flexible model specification language with a high-level algorithm language for model-generic programming, all embedded within R.  Numerous other algorithms can be envisioned for implementation with this system, such as those listed in section (\ref{rationale}) above.  

However, several important challenges remain for building out the potential of NIMBLE.  First, not all features of BUGS, or of graphical models in general, have so far been incorporated.  A particular challenge is efficient handling of stochastically indexed dependencies, such as when discrete mixture components are latent states.  This represents a dynamic graph structure and so will require a more flexible system for representing dependencies.  Second, several packages have made great use of automatic differentiation, notably ADMB/TMB and Stan.  Because the NIMBLE compiler generates C++ code, it would be possible to extend it to generate code that uses an automatic differentiation library.  Third, there is a need to include more compilable functionality in the NIMBLE DSL, such as use of R's optimization library from generated C++.  An algorithm like Laplace approximation would be most natural if optimization and derivatives are available in the DSL.  Finally, there is potential to extend the NIMBLE compiler in its own right as a useful tool for programming efficient computations from R even when there is no BUGS code involved.

The choice to embed a compilable domain-specific language within R revealed some benefits and limitations.  R's handling of code as an object facilitates processing of BUGS models and \nimbleFunction\ code.  It also allows the dynamic construction and evaluation of class-definition code for each \cd{model} and \cd{nimbleFunction} and their C++ interfaces.  And it provides many other benefits, perhaps most importantly that it allows NIMBLE to work within such a popular statistical programming environment.  On the negative side, NIMBLE needs some fundamentally different behavior than R, such as call-by-reference and functions that work by ``side effects'' (e.g., modifying an object without copying it).  Such inconsistencies make NIMBLE something of a conceptual hybrid, which could be viewed as practical and effective by some or as inelegant or confusing by others.  And for large models,  NIMBLE's compilation processing suffers from R's slow execution.  

We built upon BUGS as a model specification language because it has become so widely used, but it has limitations.  First, BUGS uses distribution notation slightly different from R, so combining BUGS and R syntaxes in the same system could be confusing.  In particular some BUGS distributions use different default parameterizations than R's distributions of the same or similar name.  Second, BUGS does not support modular model programming, such as compactly declaring common model substructures in a way that re-uses existing code.  It also does not support vectorized declarations of scalar nodes that follow the same pattern (it requires \cd{for}-loops instead).  These are extensions that could be built into NIMBLE in the future.  Other extensions, such as declaration of single multivariate nodes for vectorized calculations, were implemented almost automatically as a result of NIMBLE's design.  Third, one could envision powerful uses of programmatically generating model definitions rather than writing them in static code.  This could be done via NIMBLE's model definition system in the future.  

Other quite distinct lines of research on software for graphical models come from ``probabilistic programming'' efforts by computer scientists, such as Church \citep{goodman-etal-08} and BLOG \citep{milch-etal-06}.  Their motivations are somewhat different, and their programming style and concepts would be new to many applied statisticians.  It will be interesting to see where these two distinct motivations for similar programming language problems lead in the future.

\section{Acknowledgements}
\label{sec:acknowledgements}

We thank Jagadish Babu for contributions to an early, pre-release version of NIMBLE.  This work was supported by grant DBI-1147230 from the US National Science Foundation and by support to DT from the Berkeley Institute for Data Science.

\bibliographystyle{chicago}
\bibliography{aboutNimbleRefs.bib}

\end{document}